\documentclass[prd,aps,groupedaddress,showpacs]{revtex4}
\usepackage{epsf,epsfig,graphics}
\begin{document}


\title{Photon Production from Nonequilibrium Disoriented Chiral Condensates
\\ in a Spherical Expansion}
\author{Yeo-Yie Charng}
\email[]{charng@phys.sinica.edu.tw}
\author{Kin-Wang Ng}
\email[]{nkw@phys.sinica.edu.tw}
\affiliation{
Institute of Physics, Academia Sinica, Taipei, Taiwan, R.O.C.  }

\author{ Chi-Yong Lin}
\email[]{ lcyong@mail.ndhu.edu.tw}
\author{ Da-Shin Lee}
\email[]{ dslee@mail.ndhu.edu.tw}
\affiliation{
Department of Physics, National Dong Hwa University,
Hua-Lien, Taiwan, R.O.C.
}

\date{August 2002}
\vskip 2cm
\begin{abstract}
We study the production of photons through the non-equilibrium relaxation
of a disoriented chiral condensate formed in the expanding hot central region
in ultra-relativistic heavy-ion collisions. It is found that the expansion
smoothes out the resonances in the process of parametric amplification such that
the non-equilibrium photons are dominant to the thermal photons over the
range 0.2-2 GeV. We propose that to search for non-equilibrium
photons in the direct photon measurements of heavy-ion collisions can be a
potential test of the formation of disoriented chiral condensates.
\end{abstract}


\maketitle

In relativistic heavy ion collisions, the highly Lorentz contracted
nuclei essentially pass through each other, leaving behind a hot plasma
in the central rapidity region with large energy  density corresponding to
temperature above $200 ~{\rm MeV}$ where the chiral symmetry is restored.
This plasma then cools down via rapid hydrodynamic expansion through the
chiral phase transition during which the long-wavelength fluctuations become
unstable and grow due to the spinoidal
instabilities~\cite{wilczek,cooper,boydcc}.
The growth of these unstable modes results in the formation of
disoriented chiral condensates (DCCs) (see also Refs.~\cite{bjorken1,dcc}).
The DCCs are the correlated regions of space-time where
the chiral order parameter of QCD is chirally rotated from its usual
orientation in isospin space.
Subsequent relaxation of the DCCs to the true QCD vacuum is
expected to
radiate copious soft pions which could be a potential experimental signature
of the chiral phase transition observable in the
ultra-relativistic  heavy ion collisions at RHIC and
LHC  although
such a signature has been looked for and not seen in the lower
energy heavy ion collisions at the CERN SPS~\cite{agg}.
However, since these emitted pions will undergo the strong final interaction,
the signal may be severely masked and become indistinguishable from the
background. It then becomes important to study other possible signatures
of DCCs that would be less affected by the final state interaction.
Electromagnetic probe such as photon and lepton with longer mean free path
in the medium serves as a good candidate and can reveal more detailed
non-equilibrium information on the DCCs
with minimal distortion~\cite{wang,boy1,boy2}.

Minakawa and Muller \cite{muller} have suggested
that the presence of strong electromagnetic
fields in relativistic heavy ion collisions induces  a
quasi-instantaneous ``kick'' to the field configuration along the
$\pi^0$ direction such that it is plausible  that the chiral order parameter
in the DCC domains, if formed, will acquire a component in the direction
of the neutral pion.
The production of photons through the non-equilibrium relaxation of a DCC
within which the chiral order parameter initially has a non-vanishing
expectation value along the $\pi^{0}$ direction and subsequently oscillates
around the minimum of the effective potential has been considered.
In Ref.~\cite{boy2}, Boyanovsky et al. have extensively studied the
photon production from the low energy coupling of the neutral pion to photon
via the ${\rm U}_{\rm A} (1)$ anomalous vertex.
They have found that for large initial amplitudes of the $\pi^0$ field
photon production is enhanced by parametric amplification.
These processes are non-perturbative with a large contribution during
the non-equilibrium stages of the evolution and result in a distinct
distribution of the produced photons. Later, the authors in Ref.~\cite{leeng}
have taken into account another dominant contribution
that also involves the dynamics of $\pi^{0}$ due to the decay of the
vector meson through the electromagnetic vertex.  Although  the corresponding
dimensionless effective coupling  involving  the vector meson is quite
perturbatively small, for the large
amplitude oscillations of the $\pi^{0}$ mean field, the
contribution to the photon production  is of the same order of
magnitude as the anomalous interaction. However, they have ignored the
hydrodynamical expansion and adopted the simple ``quench'' phase transition
from an initial thermodynamic equilibrium state at a temperature ($T_i$)
higher than the critical temperature ($T_c$) for the chiral phase transition
cooled instantaneously to zero temperature,
which has been widely used in the study of
non-equilibrium phenomena of DCCs~\cite{boydcc,wang,boy1,boy2}.

In this paper, we will consider the effect of the hydrodynamical expansion
of the plasma to the production of photons from the non-equilibrium relaxation
of a DCC.  Although the ``quench'' scenario can capture the qualitative
features of this non-equilibrium problem, we will see that the dynamics of
parametric amplification as well as spinoidal instabilities is considerably
modified by the plasma expansion. In fact, our  study of the photon production
based upon  this  strongly out-of-equilibrium scenario  bears  analogy to
the recent studies in the context of the  off-shell  photon production from
the  out-of-equilibrium  expanding quark-gluon plasma with a finite life 
time~\cite{wbn}.
To make a comparison, we also compute the thermal equilibrium photon emission
from the  hot quark-gluon plasma as well as  the hadronic matter following an 
adiabatic quark-gluon phase transition
 by convoluting the equilibrium photon production rates
at finite temperature with the expansion dynamics.

In ultra-relativistic heavy-ion collisions, it is expected that
the rapidity density of the particles produced in the hot central
region has a plateau. This implies an approximate Lorentz boost
invariance along the longitudinal direction in the evolution of
the hot plasma in the central region~\cite{bjor}.
However, at late times following  the heavy nuclei collisions, a transverse
flow can be generated due to the multi-scattering between the
produced particles, as such the expansion becomes three dimensional~\cite{soll}.
Here we will simply assume that the hydrodynamical flow
is spherically symmetric, and that the boost is along the radial
direction. We will see that this assumption greatly simplifies the treatment
of the photon field in the expanding space-time. We consider this as the first
simplest attempt to tackle the problem.

The natural coordinates for spherical boost invariant
hydrodynamical flow are the  proper time $\tau$ and the space-time
rapidity $\eta$ defined as
\begin{equation}
\tau\equiv (t^2-r^2)^{1\over2},\quad
\eta\equiv {1\over2}\ln\left(\frac{t+r}{t-r}\right),
\end{equation}
where
 $(t,{\vec r})$ are the coordinates in the laboratory,
\begin{equation}
t=\tau\cosh\eta,\quad r=\tau\sinh\eta.
\label{transf}
\end{equation}
The ranges of these coordinates are set to be  $0 \le \tau <
\infty$ and $0 \le \eta < \infty$, restricted  to the forward
light cone. In terms of spherical coordinates, the Minkowski line
element is given by
\begin{eqnarray}
ds^2&=&dt^2-d{\vec r}^2 \label{lab} \\
    &=&d\tau^2- \tau^2(d\eta^2 + \sinh^2\eta\,d\theta^2
     + \sinh^2\eta\,\sin^2\theta\,d\phi^2).
\label{open}
\end{eqnarray}

Eq.~(\ref{open}) is the Robertson-Walker metric for an open
expanding universe with the scale factor
$R(\tau)=\tau$~\cite{wein}. For $\eta < 1$, the open
metric~(\ref{open}) can be approximated by the flat
Robertson-Walker metric,
\begin{eqnarray}
ds^2&=& d\tau^2- \tau^2(d\eta^2 + \eta^2\,d\theta^2
     + \eta^2\,\sin^2\theta\,d\phi^2), \nonumber \\
    &=& d\tau^2- \tau^2 d{\vec \eta}^2.
\label{flat}
\end{eqnarray}
This flat metric will serve as the background comoving frame under
which to study non-equilibrium photon production from DCC domains.
This is a good first-order approximation since the boost
invariance is in anyway applied only for small values of $\eta$.
Furthermore, $\eta < 1$ corresponds to the space-time with $r/t <
0.76$ which has covered a major portion of the forward light cone.

The relevant phenomenological effective action in a general
expanding space-time is given by
\begin{equation}
S=\int d^4x {\sqrt g} \left( L_\sigma+ L_A + L_{\pi^0 A}\right),
\label{action}
\end{equation}

where

\begin{eqnarray}
L_\sigma &=& -{1\over2}g^{\mu\nu}\partial_\mu {\vec\Phi}\cdot \partial_\nu
             {\vec\Phi} + {1\over2} \frac{M_\sigma^2}{2}
             {\vec\Phi}\cdot {\vec\Phi}
             - \lambda \left({\vec\Phi}\cdot{\vec\Phi}\right)^2 + h\sigma, \label{Lpi}\\
L_A &=& -{1\over4} g^{\alpha\mu} g^{\beta\nu} F_{\alpha\beta} F_{\mu\nu}, \label{La}\\
L_{\pi^0 A} &=& \frac{1}{\sqrt g}\frac{e^2}{32\pi^2}\frac{\pi^0}{f_\pi}
             \epsilon^{\alpha\beta\mu\nu} F_{\alpha\beta} F_{\mu\nu}
             +\frac{1}{(\sqrt g)^2} \frac{e^2\lambda_V^2}{8m_\pi^2 m_V^2}
             \epsilon^{\mu\nu\lambda\sigma}\epsilon^{\alpha\beta\gamma\delta}
             g_{\sigma\delta} \partial_\lambda \pi^0 \partial_\gamma \pi^0
             F_{\mu\nu}F_{\alpha\beta}, \label{effpia}
\end{eqnarray}
where $F_{\mu\nu}=\partial_\mu A_\nu - \partial_\nu A_\mu$ is the photon field,
and ${\vec\Phi}=(\sigma,\pi^0,{\vec\pi})$ is an $O(N+1)$ vector of scalar fields
with
 $\vec\pi=(\pi^1,\pi^2,...,\pi^{N-1})$ representing the $N-1$
 pions.
Note that the signature is $(-+++)$, and we have added $1/{\sqrt g}$ to each
$\epsilon^{\alpha\beta\mu\nu}$ because it is a tensor density of
weight $-1$~\cite{wein}. The phenomenological parameters in the
effective Lagrangian above can be determined by the low-energy
 pion
physics as follows:
\begin{eqnarray}
&& m_{\sigma} \approx 600 ~{\rm MeV}, ~~ f_{\pi} \approx 93 ~{\rm
MeV}, ~~ {\lambda} \approx 4.5, ~~ T_c \approx 200 ~{\rm MeV},
\nonumber \\ && h \approx (120~ {\rm MeV})^3,~~ m_{V} \approx 782
~{\rm MeV},~~ \lambda_{V} \approx 0.36, \label{parameter}
\end{eqnarray}
where $V$ is identified as the $\omega$ meson, and the coupling
$\lambda_{V}$ is  obtained  from the $ \omega \rightarrow \pi^0
\gamma$ decay width~\cite{davi}. This effective action has been
obtained by two of us~\cite{leeng} in the study of out-of-equilibrium photon
production from DCC domains by taking the rapid ''quench'' phase
transition scenario. Here
 we will consider the similar  non-equilibrium
phenomena taking
account of the hydrodynamical expansion.
It must be noticed that this is an effective field theory  with
an ultraviolet
 momentum cutoff of the order of $\Lambda \approx 1~$GeV.
The effective vertices obtained  are  in terms of the perturbative
theory without involving in-medium modifications. In-medium
effects will enter only  through the non-equilibrium Green's
functions for the meson fields as we will see below. To obtain
these effective vertices including the strongly out of equilibrium
effects, one should  integrate out the quark fields and
the vector meson in the context of the fully non-equilibrium
formalism.  It  deserves to tackle in the near future.

It is well known that the minimal coupling of photons to
the metric background is conformally invariant. Thus,
in the conformally flat metric~(\ref{flat}), it is convenient to
work with the conformal time defined by
\begin{equation}
du\equiv \tau_i  \frac{d\tau}{\tau},
\end{equation}
where $\tau_i$ is the initial proper time after which we expect
that the quark-gluon plasma is formed and thermalized. Integration
gives
\begin{equation}
u= \tau_i \ln\left({\tau\over \tau_i}\right).
\end{equation}
Rewrite the flat metric~(\ref{flat}) as
\begin{equation}
ds^2= -g_{\mu\nu} dx^\mu dx^\nu = a^2(u) (du^2-d{\vec x}^2),
\end{equation}
where $a(u)=\tau/\tau_i=e^{u/\tau_i}$ and $d{\vec x}=\tau_i d{\vec
\eta}$.
Writing explicitly in terms
 of the comoving coordinates and defining
${\vec\Phi}={
\vec\Phi_{a}}/a$, the action~(\ref{action}) becomes
\begin{equation}
S= \int du~d^3{\vec x}~~{\cal L}
 = \int du~d^3{\vec x}~~
   \left({\cal L}_\sigma+ {\cal L}_A + {\cal L}_{\pi^0 A}\right),
\end{equation}
where
\begin{eqnarray}
{\cal L}_\sigma &=&-{1\over2}\eta^{\mu\nu}\partial_\mu
            {\vec\Phi_{a}}
             \cdot \partial_\nu{ \vec\Phi_{a}} + {1\over2}
             \left[ {1\over2} a^2 M_\sigma^2 + {1\over a}
             \frac{d^2 a}{du^2} \right]
             { \vec\Phi_{a}}\cdot { \vec\Phi_{a}}
             - \lambda \left({ \vec\Phi_{a}}\cdot{ \vec\Phi_{a}}\right)^2 +
              a^3 h\sigma_a, \\
{\cal L}_A &=& -{1\over4} \eta^{\alpha\mu} \eta^{\beta\nu}
            F_{\alpha\beta} F_{\mu\nu}, \\
{\cal L}_{\pi^0 A} &=& \frac{e^2}{32\pi^2}\frac{\pi^0_a}{af_\pi}
             \epsilon^{\alpha\beta\mu\nu} F_{\alpha\beta} F_{\mu\nu}
             +\frac{1}{a^2} \frac{e^2\lambda_V^2}{8m_\pi^2 m_V^2}
             \epsilon^{\mu\nu\lambda\sigma}\epsilon^{\alpha\beta\gamma\delta}
             \eta_{\sigma\delta}
             \partial_\lambda \left(\frac{\pi^0_a}{a}\right)
             \partial_\gamma \left(\frac{\pi^0_a}{a}\right)
             F_{\mu\nu}F_{\alpha\beta}, 
\end{eqnarray}
where ${ \vec\Phi_{a}}=(\sigma_a,\pi^0_a,{\vec\pi}_a)$ and
$\eta^{\mu\nu}$ is the Minkowski metric. In terms of the conformal
time, the effective action now has analogy with the effective
action in Minkowski spacetime with the time dependent mass term
and interactions.

The  dynamics of the non-equilibrium expectation values as well as
correlation functions of quantum fields can be obtained by
implementing the Schwinger-Keldysh closed-time-path formulation of
non-equilibrium quantum field theory. This formulation is used  to
describe the evolution of an initially prepared density matrix,
and requires a path integral defined along a closed time contour.
This techniques have successfully been  employed elsewhere within
many different contexts and we refer the readers to the literature
for details~\cite{boyneq}. Here we first shift $\sigma_a$ and $\pi^0_a$ by
their
expectation values with respect to  an initial  non-equilibrium
states:

\begin{eqnarray}
&&\sigma_a({\vec x},u) = \phi_a(u) + \chi_a({\vec x},u),\quad\quad
  \langle\sigma_a({\vec x},u)\rangle = \phi_a(u), \\
&&\pi^0_a({\vec x},u) = \zeta_a(u) + \psi_a({\vec x},u),\quad\quad
  \langle\pi^0_a({\vec x},u)\rangle = \zeta_a(u),
\end{eqnarray}
with the tadpole conditions:
\begin{equation}
\langle\chi_a({\vec x},u)\rangle=0,\quad
\langle\psi_a({\vec x},u)\rangle=0,\quad
\langle{\vec \pi}_a({\vec x},u)\rangle=0.
\label{tad}
\end{equation}

One can impose these tadpole conditions  to all orders in the
corresponding expansion to derive the non-equilibrium evolution
equations of field expectation values. In addition, the large-$N$
approximation will be implemented below (Eqs.~(2.7)-(2.9)
of Ref.~\cite{boy2})  to
obtain the factorized Lagrangian which provides a
non-perturbative framework to  self-consistently incorporate
quantum fluctuation effects from the strong $\sigma-\pi$
interactions. As for the issue of the gauge choice, we adopt here
the Coulomb gauge which involves only the physical degrees of freedom
without any redundant fields to study  non-equilibrium photon
production from the oscillating mean fields within the DCC domains
in a spherical expansion.

Using the tadpole conditions~(\ref{tad}), the large-$N$ equations of
motion for the mean fields involving  the  perturbatively electromagnetic
corrections  can be obtained as follows:
\begin{eqnarray}
&& \left[ \frac{d^2}{du^2} +m_a^2(u) +4\lambda\phi_a^2(u)
+4\lambda\zeta_a^2(u) +4\lambda\langle{\vec \pi}_a^2\rangle(u)
 \right] \phi_a - a^3 h=0, \nonumber \\
&&\left[ \frac{d^2}{du^2} +m_a^2(u) +4\lambda\phi_a^2(u)
+4\lambda\zeta_a^2(u) +4\lambda\langle{\vec
\pi}_a^2\rangle(u)\right] \zeta_a -\frac{e^2}{32\pi^2 f_\pi}
{1\over a} \langle F {\tilde F} \rangle \nonumber \\
&&+\frac{e^2\lambda_V^2}{4m_\pi^2 m_V^2} {1\over a} {d\over
du}\left[{1\over a^2}{d\over du}
\left(\frac{\zeta_a}{a}\right)\right] \epsilon^{\mu\nu
0\sigma}\epsilon^{\alpha\beta 0\delta} \eta_{\sigma\delta} \langle
F_{\mu\nu}F_{\alpha\beta}\rangle \nonumber \\
&&+\frac{e^2\lambda_V^2}{4m_\pi^2 m_V^2} {1\over a^3}{d\over du}
\left(\frac{\zeta_a}{a}\right) \epsilon^{\mu\nu
0\sigma}\epsilon^{\alpha\beta \gamma\delta}\eta_{\sigma\delta}
\langle \partial_{\gamma} F_{\mu\nu}F_{\alpha\beta}\rangle=0.
\label{eommf}
\end{eqnarray}
We then decompose the fields ${\vec \pi}_a$ and ${\vec A}_T$ into
their Fourier mode functions $U_{a\vec k}(u)$ and $V_{\lambda \vec
k}(u)$ respectively,
\begin{eqnarray}
{\vec \pi}_a(\vec x,u)
&=& \int \frac{d^3{\vec k}}{\sqrt{2(2\pi)^3\omega_{{\vec\pi}_a {\vec k},i}}}
\left[{\vec a}_{\vec k} U_{a\vec k}(u) e^{i\vec k\cdot \vec x} +
{\rm h.c.}\right], \\
\vec A_T(\vec x,u)
&=& \int \frac{d^3{\vec k}}{\sqrt{2(2\pi)^3 \omega_{A{\vec k},i}}}
    {\vec A_T}( {\vec k},u) \nonumber \\
&=& \int \frac{d^3{\vec k}}{\sqrt{2(2\pi)^3 \omega_{A{\vec k},i}}}
  \left\{ \left[b_{+ \vec k} V_{1 \vec k}(u) \vec \epsilon_{+ \vec k}
+ b_{- \vec k} V_{2 \vec k}(u) \vec \epsilon_{- \vec k} \right]
 e^{i\vec k\cdot \vec x} + {\rm h.c.} \right\},
\end{eqnarray}
where ${\vec a}_{\vec k}$ and $b_{\pm \vec k}$ are destruction
operators, and ${\vec \epsilon}_{\pm \vec k}$ are circular
polarization unit vectors. The frequencies $\omega_{{\vec
\pi}_a\vec k,i}$ and $\omega_{A\vec k,i}$ can be determined from
the initial states and will be specified below. The  mode
equations for the photons as well as the pions can be read off
from the quadratic part of the effective factorized
Lagrangian as
\begin{eqnarray}
&&\left[ \frac{d^2}{du^2} +k^2 +m_a^2(u) +4\lambda\phi_a^2(u)
+4\lambda\zeta_a^2(u) + 4\lambda\langle{\vec
\pi}_a^2\rangle(u)\right]U_{ak}=0, \label{ugap} \\
&&\left\{ \frac{d^2}{du^2} + \left[1-\frac{e^2\lambda_V^2}{m_\pi^2
m_V^2} \frac{{\dot\zeta}^2}{a^2} \right] k^2 - \frac{e^2}{2\pi^2
f_\pi} {\dot \zeta} k \right\} V_{1 k}=0, \label{v1gap} \\
&&\left\{ \frac{d^2}{du^2} + \left[1-\frac{e^2\lambda_V^2}{m_\pi^2
m_V^2} \frac{{\dot\zeta}^2}{a^2} \right] k^2 + \frac{e^2}{2\pi^2
f_\pi} {\dot \zeta} k \right\} V_{2 k}=0, \label{v2gap}
\end{eqnarray}
where $ m_a^2(u)= -(1/2)~a^2 M_\sigma^2 - \ddot a/a$,
$\zeta(u)=\zeta_a(u)/a$, $k=|\vec k|$, and the dot means $d/du$.
The expectation values with respective to the initial states are
given by
\begin{eqnarray}
\langle{\vec \pi}_a^2\rangle(u) &=& (N-1) \int^{\Lambda a(u)}
\frac{d^3{\vec k}}{2(2\pi)^3\omega_{{\vec \pi}_a k}}
\left[ |U_{ak}(u)|^2 \right]
\coth \left[ \frac{\omega_{{\vec \pi}_a k}}{2 T_i} \right]  , \nonumber \\
\epsilon^{\alpha\beta\mu\nu} \langle F_{\alpha\beta} F_{\mu\nu}
\rangle(u) & =&   \int^{\Lambda a(u)}
\frac{d^3{\vec k}}{2(2\pi)^3\omega_{A k}} (4k) \frac{d}{du} \left[ |V_{2
k}(u)|^2- |V_{1 k}(u)|^2 \right], \nonumber \\
\epsilon^{\mu\nu 0\sigma}\epsilon^{\alpha\beta 0\delta}
\eta_{\sigma\delta} \langle F_{\mu\nu}F_{\alpha\beta}\rangle (u)
&=&
 \int^{\Lambda a(u)} \frac{d^3{\vec k}}{2(2\pi)^3\omega_{A k}} (4 k^2)
\left[ |V_{1 k}(u)|^2+ |V_{2 k}(u)|^2 \right], \nonumber \\
\epsilon^{\mu\nu 0\sigma}\epsilon^{\alpha\beta
\gamma\delta}\eta_{\sigma\delta} \langle \partial_{\gamma}
F_{\mu\nu}F_{\alpha\beta}\rangle (u) &=&
 \int^{\Lambda a(u)} \frac{d^3{\vec k}}{2(2\pi)^3\omega_{A k}} (4 k^2)
 \frac{d}{du} \left[ |V_{1 k}(u)|^2+ |V_{2 k}(u)|^2 \right], \label{qf}
\end{eqnarray}
where $\langle{\vec \pi}_a^2\rangle(u)$ is self-consistently
determined by Eq.~(\ref{ugap}). In particular, when $u=0$, it
becomes the gap equation,
\begin{equation}
\langle{\vec \pi}_a^2\rangle(0)= (N-1) \int^\Lambda\frac{d^3{\vec
k}}{2(2\pi)^3\omega_{{\vec \pi}_a k,i}}
\coth\left[\frac{\omega_{{\vec \pi}_a k,i}}{2T_i}\right].
\end{equation}
Notice  that the  photonic medium effects have been ignored in the
above expressions of the photon field expectation values due to
the fact that  the photons interact electromagnetically so that
their mean free paths are expected to be longer than the estimated
size of the  quark-gluon plasma fireball, and the produced photons
will escape freely toward the particle detector having no enough
time to build up their population in the plasma. In this conformal  frame,
the momentum cutoff we choose   depends linearly on $a$ so as to
keep the physical momentum cutoff $ \Lambda $ fixed in the
laboratory frame. As a result, the initial conditions for the mode
functions become more subtle here. Here we choose
\begin{eqnarray}
&&U_{a{k < \Lambda}}(0)=1,\;\dot U_{a {k < \Lambda}}(0)=
-i\omega_{{\vec \pi}_a k,i}, \nonumber \\
&&\omega_{{\vec\pi}_ak,i}^2 = k^2 + m_a^2(0) +4\lambda\phi_a^2(0)
+4\lambda\zeta_a^2(0) + 4\lambda\langle{\vec \pi}_a^2\rangle(0);
\nonumber \\
 &&V_{1,2 {k < \Lambda}}(0)=1,\;\dot V_{1,2 { k <
\Lambda}}(0)= -i\omega_{A k,i},\;\omega_{A k,i}=k;
\label{u0}
\\
&&U_{a { k > \Lambda}}(u_k )=1,\;\dot U_{a {k > \Lambda}}(u_k)=
-i\omega_{{\vec \pi}_a k,i}, \nonumber
\\ &&\omega_{{\vec \pi}_a k,i}^2 = k^2 + m_a^2(u_k)
+4\lambda\phi_a^2(u_k) +4\lambda\zeta_a^2(u_k) +
4\lambda\langle{\vec \pi}_a^2\rangle(u_k); \nonumber \\
&&V_{1,2{k> \Lambda}}(u_k)=1,\; \dot V_{1,2 {k > \Lambda}}(u_k)=
-i\omega_{A k,i},\;\omega_{A k,i}=k.
\label{uk}
\end{eqnarray}
Therefore, for  the momentum lying  below the momentum cutoff
$\Lambda$ at the initial conformal time, the initial condition for
the mode function can be set at  $u=0$ as shown in Eq.~(\ref{u0}).
However, for the momentum
above the cutoff momentum $\Lambda$  at $u=0$,  it will become
dynamical when $k \le \Lambda a(u)$ after $ u=u_k = \tau_i
{\rm ln}(k/\Lambda)$, and the initial condition for this mode function
must be set at that conformal time $u_k$ as in Eq.~(\ref{uk}).
The photon spectral number density at time $u$ is given by the
expectation value of the number operator for the asymptotic
photons~\cite{boy2},
\begin{eqnarray}
\frac{dN}{d^3{\vec x} d^3{\vec k}} &=& \frac{dN_+}{d^3{\vec x}
d^3{\vec k}} + \frac{dN_-}{d^3{\vec x} d^3{\vec k}} \nonumber \\
&=& \langle {\bf N}_{k}(u)\rangle \nonumber \\
&=&
{\frac{1}{2k}} \left[ \dot{\vec A_T}( {\vec k},u) \cdot\dot{\vec
A_T} ( -{\vec k},u) +k^2 {\vec A_T}( {\vec k},u) \cdot
  {\vec A_T}({-\vec k},u)\right]-1    \nonumber \\
&=& {1\over 4k^2}  \left[|\dot V_{1 k}(u)|^2+k^2 | V_{1
k}(u)|^2\right] + {1\over 4k^2}  \left[|\dot V_{2 k}(u)|^2+k^2 | V_{2
k}(u)|^2\right] - 1,
\end{eqnarray}
and the invariant photon spectral production rate is given by~\cite{boy2,leeng}
\begin{equation}
\frac{kdN}{du d^3{\vec x} d^3{\vec k}} = \frac{1}{4} \left\{
\frac{e^2\lambda_V^2}{m_\pi^2 m_V^2} \frac{{\dot\zeta}^2}{a^2} k
\frac{d}{du} \left[ |V_{1 k}|^2+ |V_{2 k}|^2 \right] +
\frac{e^2}{2\pi^2 f_\pi}
   {\dot\zeta} \frac{d}{du} \left[ |V_{1 k}|^2- |V_{2 k}|^2 \right] \right\}.
   \label{conformalrate}
\end{equation}

We now need to relate this spectral production rate to the
invariant production rate measured in the metric~(\ref{lab})
in order to obtain the photon spectrum in the laboratory
 frame.  From the coordinate
 transformations~(\ref{transf}), we find that
\begin{eqnarray}
q &=& p \cosh\eta + p_r \sinh\eta, \nonumber \\ q_\eta &=& p \tau
\sinh\eta + p_r \tau \cosh\eta , \nonumber \\ q_\theta &=&
p_\theta, \nonumber \\ q_\phi &=& p_\phi, \label{eta-r}
\end{eqnarray}
and the zero-mass condition of the photon, i.e., $g^{\mu\nu}p_\mu
p_\nu=0$,
\begin{eqnarray}
q&=& \frac{|\vec q|}{\tau} = \frac{1}{\tau}
\left(q_\eta^2+\frac{1}{\sinh^2\eta} q_\theta^2
     +\frac{1}{\sinh^2\eta \sin^2\theta} q_\phi^2 \right)^{1\over 2},
     \nonumber \\
p&=& |\vec p| = \left(p_r^2+\frac{1}{r^2} p_\theta^2
     +\frac{1}{r^2\sin^2\theta} p_\phi^2 \right)^{1\over 2},
\label{mass}
\end{eqnarray}
where $q_\mu=(q,\vec q)$ is the photon four-momentum in the
metric~(\ref{open}) and  $p_\mu=(p,\vec p)$ is that
measured in the laboratory frame.
Because of the spherical symmetry of the problem, for convenience,
we can choose $\vec p$ to be along the $z$-axis, which is also the
polar axis. This gives $p_r= p\cos\theta$, and then from
Eq.~(\ref{eta-r}),
\begin{equation}
q=p(\cosh\eta+\cos\theta\sinh\eta). \label{qp}
\end{equation}
Hence, from the momentum transformation laws (\ref{eta-r}),
one can change the momentum variables from $q_\mu$ into $p_\mu$ in
the invariant production rate making use of   the fact that the
spectral particle number density $dN$ is invariant under such
transformations. We then find that
\begin{equation}
\frac{ dN}{  d\Gamma} =\frac{q~ dN}{dq_\eta dq_\theta
dq_\phi d\tau d\eta d\theta d\phi }=  \frac{p~ dN}{d^3{\vec p}~
\tau^3 \sinh^2\eta \sin\theta d\tau d\eta d\theta d\phi },
\end{equation}
where $ d\Gamma$ stands for  the invariant phase space
element in the comoving frame.
Therefore, the  photon spectrum measured in the laboratory at the
final time $\tau_f$ can be obtained  by integrating  the invariant
spectral production  rate over the space-time history of the
hydrodynamical evolution in the comoving frame as
\begin{equation}
\frac{p~dN}{d^3{\vec p}}= \int_{\tau_i}^{\tau_f} d\tau
             \int_0^{\eta_{\rm max}} d\eta
              \int_0^\pi d\theta
               \int_0^{2\pi} d\phi~
               \tau^3 \sinh^2\eta \sin\theta
               \frac{ dN}{  d\Gamma} (\tau ,q)~,
                \label{photospectrum}
\end{equation}
where $q=p(\cosh\eta+\cos\theta\sinh\eta)$. Now, in the small
$\eta$ approximation, one can link the invariant rate in the
comoving frame to that of the conformal frame as given in
Eq.~(\ref{conformalrate}), where $k=q\tau/\tau_i$ and $\vec k=\vec
q/\tau_i$, in the following way:
\begin{equation}
\frac{ dN}{  d\Gamma}\simeq \frac{q~ dN}{ d\tau
         d^3{\vec \eta} d^3{\vec q}}
  \simeq \left( \frac{\tau_i}{\tau}
         \right)^{2} \frac{k~ dN}{ du d^3{\vec x} d^3{\vec k}}~,
\end{equation}
where $u=\tau_i {\rm ln}(\tau/\tau_i)$. Then, we can approximate
\begin{eqnarray}
\frac{p~dN}{d^3{\vec p}} &\simeq& \int_{\tau_i}^{\tau_f} d\tau
               \int_0^{\eta_{\rm max}} d\eta
                  \int_0^\pi d\theta
              \int_0^{2\pi} d\phi~
              \tau^2 \tau_i \sinh^2\eta \sin\theta
              \frac{k~dN}{ du d{\vec x} d{\vec k} } (u ,k) \nonumber \\
        &\simeq& \frac{2\pi  \tau_i^2}{p} \int_{\tau_i}^{\tau_f}
        d\tau
     \int_{\tau p e^{-\eta_{\rm max}}/\tau_i}^{\tau p e^{\eta_{\rm max}}/\tau_i} dk
     \left[ \cosh(\eta_{\rm max})- \frac{1}{2} \left( \frac{k}{p} \frac{\tau_i}{\tau} +
     \frac{p}{k} \frac{\tau}{\tau_i} \right)\right] \frac{k~dN}{du d^3{\vec x} d^3{\vec k}} (u
     ,k)~.
     \label{dccplab}
\end{eqnarray}
The final expression is obtained after changing the integration
variable from $\theta$ to $k$ as well as carrying out  the $\phi$-
and $\eta$- integrals.

In contrast with the non-equilibrium photon production from the DCC
domains, the {\it equilibrium} photon emission arising from an
expanding quark-gluon plasma (QGP)
 as well as hadronic matter can be obtained by
convoluting the equilibrium photon production
rates at finite temperature with the expansion dynamics. We
currently believe that after the ultra-relativistic heavy
collisions, in the central rapidity regime, parton-parton
multi-scatterings cause the produced quarks and gluons to reach
local thermal equilibrium at the time scale of $\tau_i$, and then
the so-called QGP with high energy/entropy density is expected
to be formed. After the formation of a QGP, the dynamics of the thermal
plasma is dominated by the hydrodynamical expansion.
The
 hydrodynamical approach is appropriate
since in this stage quarks and gluons  interact in such a dense regime
that their mean free paths are much smaller than typical
wavelengths of various collective phenomena. Subsequently, the plasma
cools adiabatically according to the equation of state of a hot
QGP down to the quark-hadron phase transition when the QGP starts
to hadronize. Assuming that the phase transition is a first order
one, quarks, gluons, and hadrons will  coexist in a mixed phase in
which the released latent heat keeps the temperature of the plasma
unchanged at the critical temperature even though
the plasma continues to expand. This mixed phase will exist until
all the matter has converted into the hadrons. The system in the
hadronic phase will continuously expand with the dynamics governed
by the equation of state of a hot hadron gas. As a result,
 the
temperature continuously  drops till  the so-called freeze-out
temperature is reached where the mean free path for all hadrons is
of the order of the size of the plasma, and after that all hadrons
cease further interactions. We would like to emphasize at this
stage that according to such a hydrodynamical picture, the
strongly out of equilibrium scenario of either supercooling or
superheating has been ignored in the sense that the entropy of the
system is assumed to be locally conserved for the whole evolving
history of the system. Therefore, photon emission under this
adiabatic expansion is in striking contrast with that from the DCC
domains associated with a strongly out of equilibrium scenario. In
the Bjorken's hydrodynamical  model, the QGP as well as hadronic
matter can be described as a fluid in local thermal equilibrium.
The energy density, pressure, entropy etc. will depend on the
proper time only, and will not depend on the space-time rapidity
under the assumption of a  boost invariance in the central
rapidity region. The dynamics of the fluid is determined by the
conservation laws of the energy-momentum, baryon number, and
entropy. In addition, an equation of state of the fluid in the
various phases can serve as an input to close this set of
dynamical equations to be solved self-consistently. Here, to be
consistent with the calculation of the non-equilibrium photon
production from DCC domains, we also assume that the underlying
hydrodynamical  expansion is of a spherical expansion with the
coordinates of the proper time $\tau$ and the space-time rapidity
$\eta$  that results from free-streaming particles with a
collective velocity $ v=r/t$ along the radial direction, where the
fluid rapidity is equal to the space-time rapidity. In the central
rapidity region which is baryon free, the conservation law of
entropy for such an adiabatic expansion reads
\begin{equation}
             s(T)\tau^3 = s(T_i) \tau_i^3.
             \label{entropyconserved}
\end{equation}
This conservation law will play an important role in
determining the cooling law as well as the duration time
scales of various phases of the evolving system from the
initial QGP phase, to the quark-hadron mixed phase, and to the final
hadronic phase.

In the QGP phase, the entropy density is dominated by the
relativistic massless gas  composed of quarks
and gluons given by
\begin{equation}
               s_Q=\frac{2\pi^2}{45} g_G T^3, \label{entropyQGP}
\end{equation}
where the degeneracy $g_G = (2 \cdot 8+ 2 \cdot 2\cdot 3\cdot 2
\cdot \frac{7}{8})$ for the two-flavor quarks  and the eight SU(3)
gluons. On the other hand, the entropy density  of  hadronic
matter is taken as
\begin{equation}
               s_H=\frac{2\pi^2}{45} g_H (T) T^3, \label{entropyH}
\end{equation}
where $ g_H (T)$ is an effective degeneracy for a hadron gas. We
introduce a temperature dependence in this effective degeneracy to
account for the medium effects from the  coupled hadron gas in the
temperature regime of our  interest. The variation of the
effective degeneracy can be parametrized as
$g_H (T) \sim 13 (T/{\rm fm}^{-1})^{\delta}$
where the generally acceptable value of $ \delta$ is
about $\delta \simeq 3.4$ of slight model-dependence
(here we adopt the Walecka model)~\cite{alam}.
We are now in a position to determine the time scale in the QGP
phase which starts from the initial formation time $\tau_i$ until
$\tau_Q$ when the temperature drops to the critical temperature
$T_c$ of the quark-hadron phase transition and a phase transition
to hadronic matter starts. Following the conservation law of
entropy together with an expression of $s_Q$ in
Eq.~(\ref{entropyQGP}), the cooling law in the QGP phase can be
obtained as  $T =(\tau_i/\tau)T_i$, and thus $\tau_Q
=(T_i/T_c)\tau_i$. However, during the quark-hadron mixed phase,
the relative proportion of QGP to hadronic matter must vary with time.
Since the temperature of the quark-hadron mixed state
 remains constant at $T= T_c$ even though the plasma continuously
 expands, the  entropy density of the mixed state is equal to the sum of
the  entropy densities of QGP and hadronic matter at $T=T_c$,
 $ s^c_Q $ and $ s^c_H$, respectively
 weighted by their fractions, $ f_Q(\tau)$ and  $f_H (\tau)$:
\begin{equation}
         s_{mix}(\tau) =f_Q (\tau) s^c_Q + f_H (\tau) s^c_H,
         \label{entropyMix}
\end{equation}
where  $ f_Q(\tau)+ f_H (\tau)=1$.
Initially at $\tau=\tau_Q$,
the phase consists entirely of QGP, i.e. $f_Q (\tau_Q)=1$ ,
and at the end, entirely of hadronic matter at $\tau_H$, i.e.
$f_H(\tau_H)=1$. Thus, the life time of the mixed state is $\tau_H -\tau_Q$.
$f_Q(\tau)$, $f_H (\tau)$, and $\tau_H$ are to be
determined from the entropy conservation law~(\ref{entropyconserved})
which presumably also holds in the mixed state.
Then, from Eqs.~(\ref {entropyMix}) and (\ref{entropyconserved}), we have
\begin{equation}
   \left( f_Q (\tau) s^c_Q + [1-f_Q (\tau)] s^c_H \right) \tau^3
   =s^c_Q \tau_Q^3.
\end{equation}
Hence, we obtain
\begin{equation}
    f_Q (\tau)=\frac{ s_Q^c \left( \frac{\tau_Q}{\tau} \right)^3
    -s_H^c}{s_Q^c-s_H^c}~,~~~~~~~ f_H(\tau)=1-f_Q(\tau).
\end{equation}
Therefore,  the mixed phase ends at $\tau_H = (s_Q^c/s_H^c )^{1/3}
\tau_Q$ determined from $ f_Q(\tau_H)=0$ above. As for the
hadronic phase, substituting the associated  entropy density of a
hot hadron gas~(\ref{entropyH}) into the entropy conservation
law~(\ref{entropyconserved}) can lead to the cooling law in this
phase given by $ T=T_c (\tau_H/\tau)^{3/(\delta+3)}$ from which the
freeze-out time can be obtained as $ \tau_f = \tau_H
(T_c/T_f)^{(\delta+3)/3}$.

 After having obtained all
relevant duration time scales in various phases from the QGP
phase  to the hadronic phase, we now turn into the discussions of
the photon production processes from the QGP and the
hadronic phase. The calculation of the thermal photon
production rates in the high  energy regime $(E
>>T)$ from the
QGP at one-loop order in the hot thermal loop (HTL) approximation
has been obtained by Kapusta {\it et al.} to account for the real
photon production from the annihilation of a quark-antiquark pair
into a photon and a gluon ($q {\bar{q}} \rightarrow g \gamma$) as
well as the absorption of a gluon by a quark (antiquark) emitting
a photon ($q (\bar{q}) g \rightarrow q (\bar{q})  \gamma$),
similar to Compton scattering in QED~\cite{kapu}. However, the
recent study by Aurenche {\it et al.} has shown that the two-loop
contributions to the photon production under the HTL
approximation from quark (antiquark) bremsstrahlung ( $q q(g)
\rightarrow q q(g)\gamma)$, and  quark-antiquark annihilation with
scattering ( $ q \bar{q} q(g) \rightarrow q(g) \gamma$) are of the
same order of magnitude as that in one-loop order~\cite{aure}.
Their contributions to the photon production rates prove to be
dominant at high photon energies. Later, we will adopt this
two-loop result (Eqs.~(1,2,3) of Ref.~\cite{steffen}) to calculate the
equilibrium photon emission during the phases that involve the
QGP. As for the photon production processes from the hadronic phase, it
is known that the dominant processes in the energy regime of our
interest are those of the reactions $\pi\pi \rightarrow
\rho\gamma, \pi\rho \rightarrow \pi\gamma$ as well as the decays
$\omega \rightarrow \pi\gamma, \rho \rightarrow
\pi\pi\gamma$~\cite{nadeau}. The reaction for producing
photons that involves the intermediary axial vector meson $a_1$ is
also important,  for example, $\pi\rho \rightarrow a_1 \rightarrow
\pi\gamma$~\cite{xiong}. All of these hadronic processes  will be
taken into account in the calculation of the equilibrium photon
emission from the thermal hadrons.
 A point to be stressed is that we consider the photon
production processes from the low-mass hadrons only
while the higher mass resonances would be Boltzmann suppressed.
 We then use the invariant photon production rates
  either from the QGP or from the hadrons combining with the relevant
cooling laws as well as  their  expansion dynamics to numerically
obtain  the total equilibrium  photon emission for the whole evolution
of the system. Notice that the equilibrium photon production
invariant  rates obtained in the literature
~\cite{alam,peit} in the rest frame of the emitting
matter  can be expressed in terms of the comoving frame of a
thermal bath  with temperature being a function of the proper time
$\tau$ under the adiabatic expansion. We can denote those
invariant  rates from the QGP and the
hadron gas in the comoving frame respectively by
$dN/d\Gamma\mid_{Q}$  and $dN/d\Gamma\mid_{H}$
that both depend on the photon four-momentum $q_\mu$ as
well as the temperature $T(\tau)$ with $\tau$ dependence following
their respective cooling laws. Finally, the equilibrium photon
emission starting from the formation time $\tau_i$ down to the
freeze-out time $\tau_f$ in the laboratory frame can be obtained by
a Lorentz boost from the comoving frame to the center of mass
frame. Thus it becomes with Eq.~(\ref{photospectrum})
\begin{eqnarray}
&& \frac{p~dN}{d^3{\vec p}} = \int_{\tau_i}^{\tau_Q} d\tau
             \int_0^{\eta_{\rm max}} d\eta
              \int_0^\pi d\theta
               \int_0^{2\pi} d\phi~
               \tau^3 \sinh^2\eta \sin\theta
               \frac{ dN}{  d\Gamma} \mid_{Q}
                \left(q,T=T_i \frac{\tau_i}{\tau} \right) \nonumber \\
               &+& \int_{\tau_Q}^{\tau_H} d\tau
             \int_0^{\eta_{\rm max}} d\eta
              \int_0^\pi d\theta
               \int_0^{2\pi} d\phi~
               \tau^3 \sinh^2\eta \sin\theta
              \left( f_Q(\tau) \frac{ dN}{  d\Gamma} \mid_{Q} (q, T_c)
              + f_H (\tau) \frac{ dN}{  d\Gamma} \mid_{H} (q, T_c) \right) \nonumber  \\
           &+& \int_{\tau_H}^{\tau_f} d\tau
             \int_0^{\eta_{\rm max}} d\eta
              \int_0^\pi d\theta
               \int_0^{2\pi} d\phi~
               \tau^3 \sinh^2\eta \sin\theta
               \frac{ dN}{  d\Gamma} \mid_{H}
               \left( q, T=T_c \left(\frac{\tau_H}{\tau}\right)^{\frac{3}{\delta
               +3}}\right), \nonumber \\
               \label{tplab}
\end{eqnarray}
where again  $q=p(\cosh\eta+\cos\theta\sinh\eta)$ in Eq.~(\ref{qp}).

\begin{figure}[htbp]
\begin{center}
\leavevmode
\epsfxsize 5in
\epsfbox{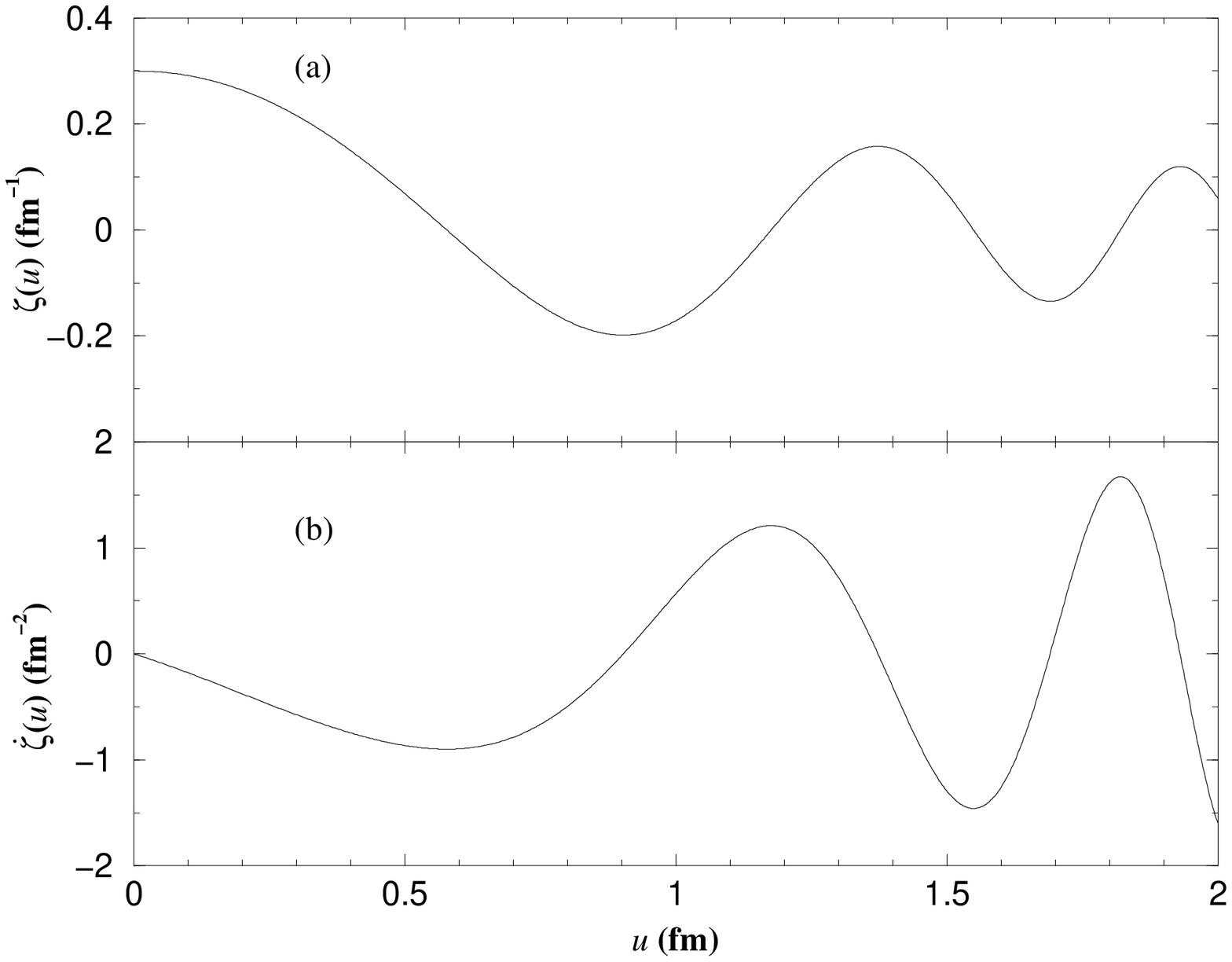}
\epsfxsize 5in
\epsfbox{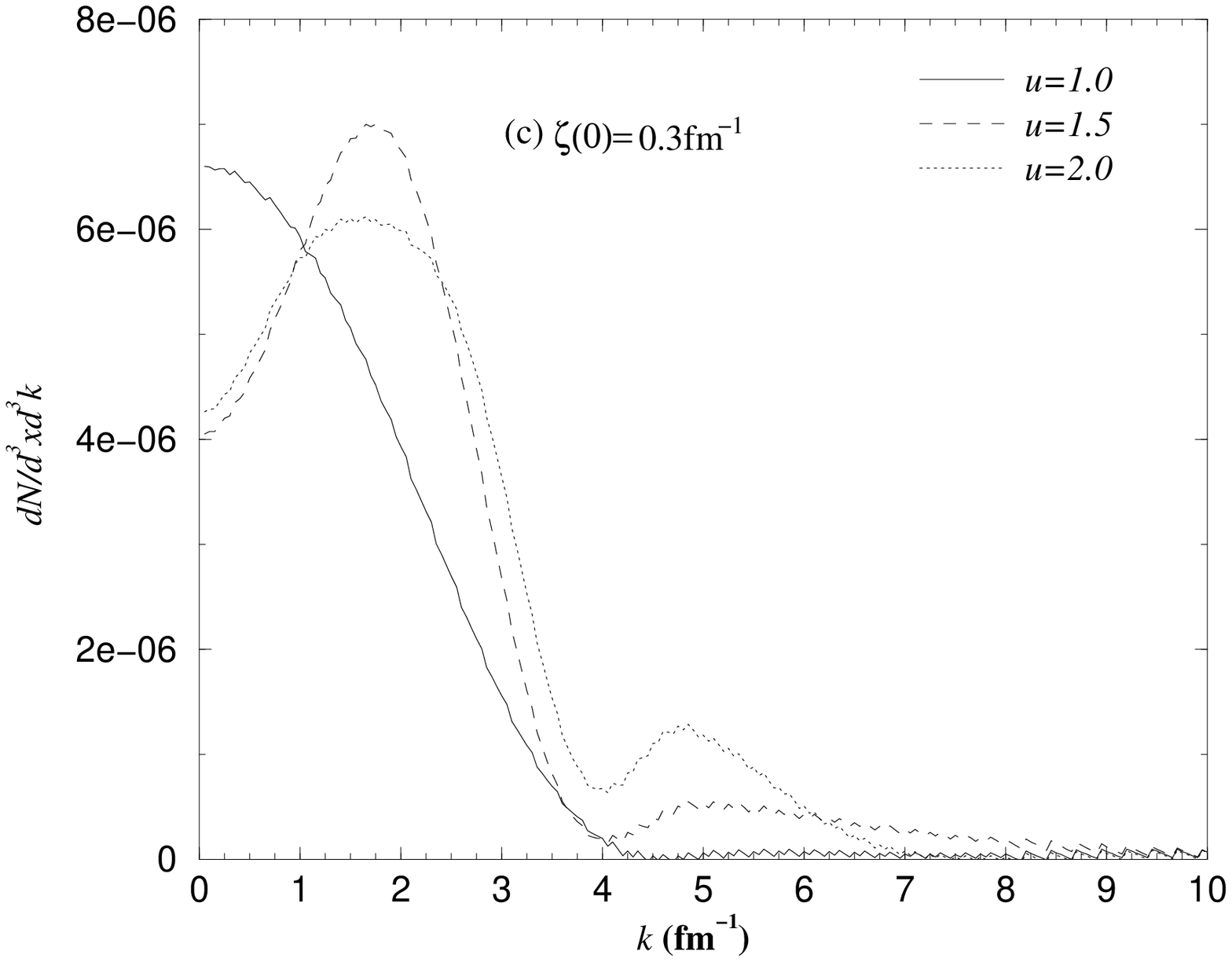}
\caption{($a$) Temporal evolution of the $\pi^0$ mean field $\zeta(u)$.
($b$) Evolution of the derivative $\dot{\zeta}(u)$.
($c$) Photon spectral number density ${dN}/{d^3xd^3k}$ for
$\zeta(0)= 0.3~ \rm{fm}^{-1}$ at different times,
where $u$ is in units of $\rm{fm}$. }
\label{fig1}
\end{center}
\end{figure}

We now perform the numerical study.  We first solve the coupled
differential equations Eq.~(\ref{eommf}),
Eqs.~(\ref{ugap}-\ref{v2gap}), and the field expectation values
Eq.~(\ref{qf}) self-consistently with the initial conditions
$\zeta(0)= 0.3~\rm{fm}^{-1}$ and $\dot{\zeta}(0)
=\phi(0)=\dot{\phi}(0)=0$ in the conformal time frame. The initial
temperature $T_i$ at the QGP formation time $\tau_i=
1.0~\rm{fm}^{-1}$ that corresponds to the initial conformal time
$u=0$ is chosen to be $T_i =1.0~\rm{fm}^{-1}$.
 We set $N=3$ for 3 pions.  In Figs.~1(a) and 1(b), we
show that $\zeta(u)$ evolves with strong damping,
while the amplitude of the field derivative $\dot{\zeta} (u)$ grows.
The strong damping is due to the back-reaction effects from the creation of
excitations of the quantum fluctuations via parametric
amplification. This translates into photon production as we will
discuss later. In fact, the growing $\dot{\zeta} (u)$ makes photon
production become more effective.  The spectra of emitted photons
in the conformal time frame are depicted for $ u=1$, $1.5$, and $2~\rm{fm}^{-1}$
respectively in Fig.~1(c). They feature prominent
peaks. Notice that the peaks are located at $k \simeq
\omega_{\dot{\zeta}}/2$ and $k\simeq\omega_{\dot{\zeta}}$, where
$\omega_{\dot{\zeta}}$ is the time-averaged oscillation frequency
of $\dot{\zeta}$ which can be read off from Fig.~1(b). Since
$\omega_{\dot{\zeta}}$ increases with time, both of the peaks move
towards the higher momenta as we can see from Fig.~1(c).
Therefore, it is evident that the photon production is due to
parametric amplification. Figs.~2(a), 2(b), and 2(c) are shown
for a large initial amplitude $\zeta(0)= 1.0~ \rm{fm}^{-1}$ and
$\dot{\zeta}(0) =\phi(0)=\dot{\phi}(0)=0$. This is clearly the
non-perturbative effects driven by the parametric amplification which exhibits
the features of unstable bands leading to the growth of the
fluctuation modes. The growth of the fluctuations turns into the
profuse photon production in the modes within the unstable bands.
It is not surprising that with a large initial field
expectation value, the peaks in the photon spectrum (Fig.2(c)) are
about an order of magnitude larger than the previous case.

\begin{figure}[htbp]
\begin{center}
\leavevmode
\epsfxsize 5in
\epsfbox{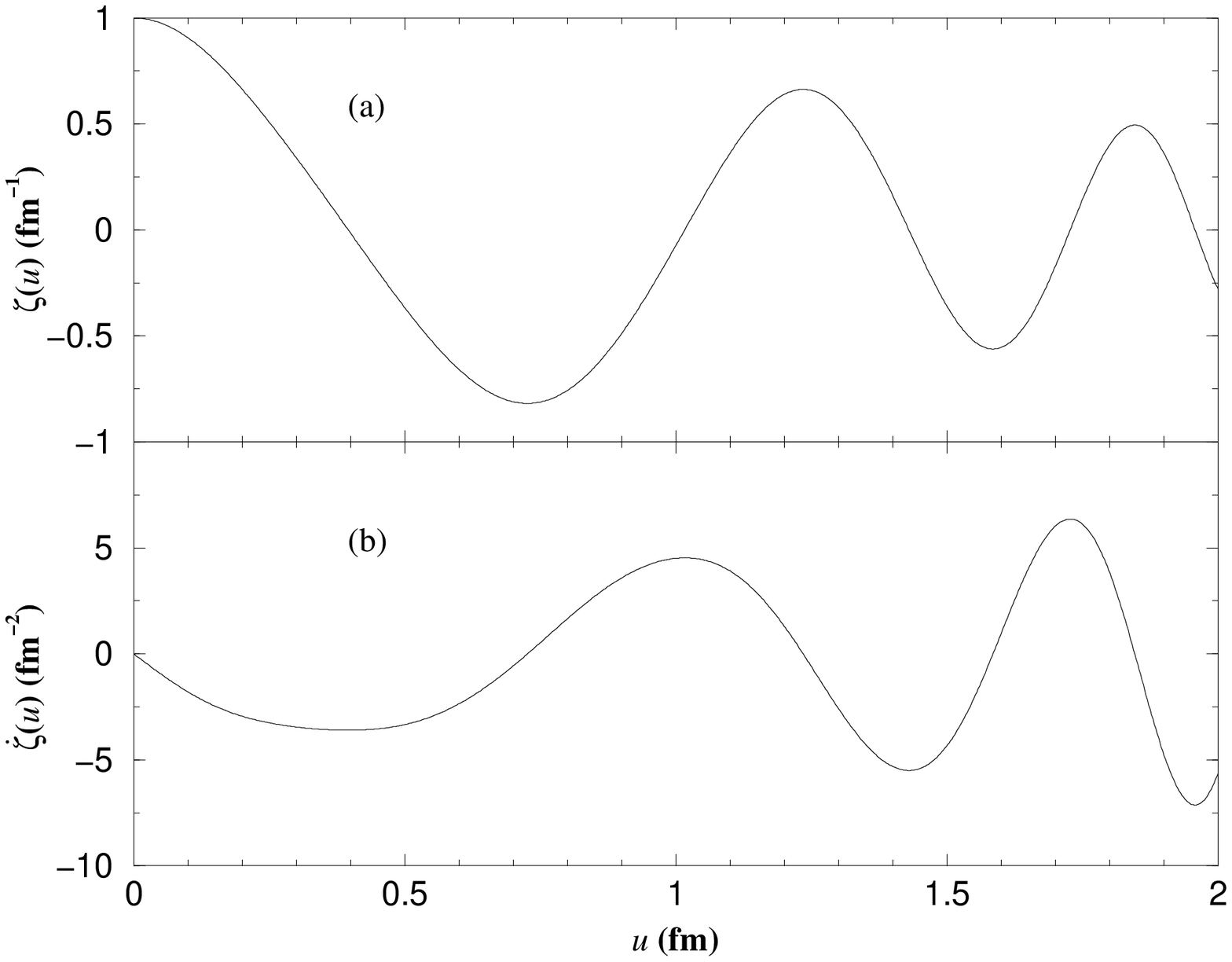}
\epsfxsize 5in
\epsfbox{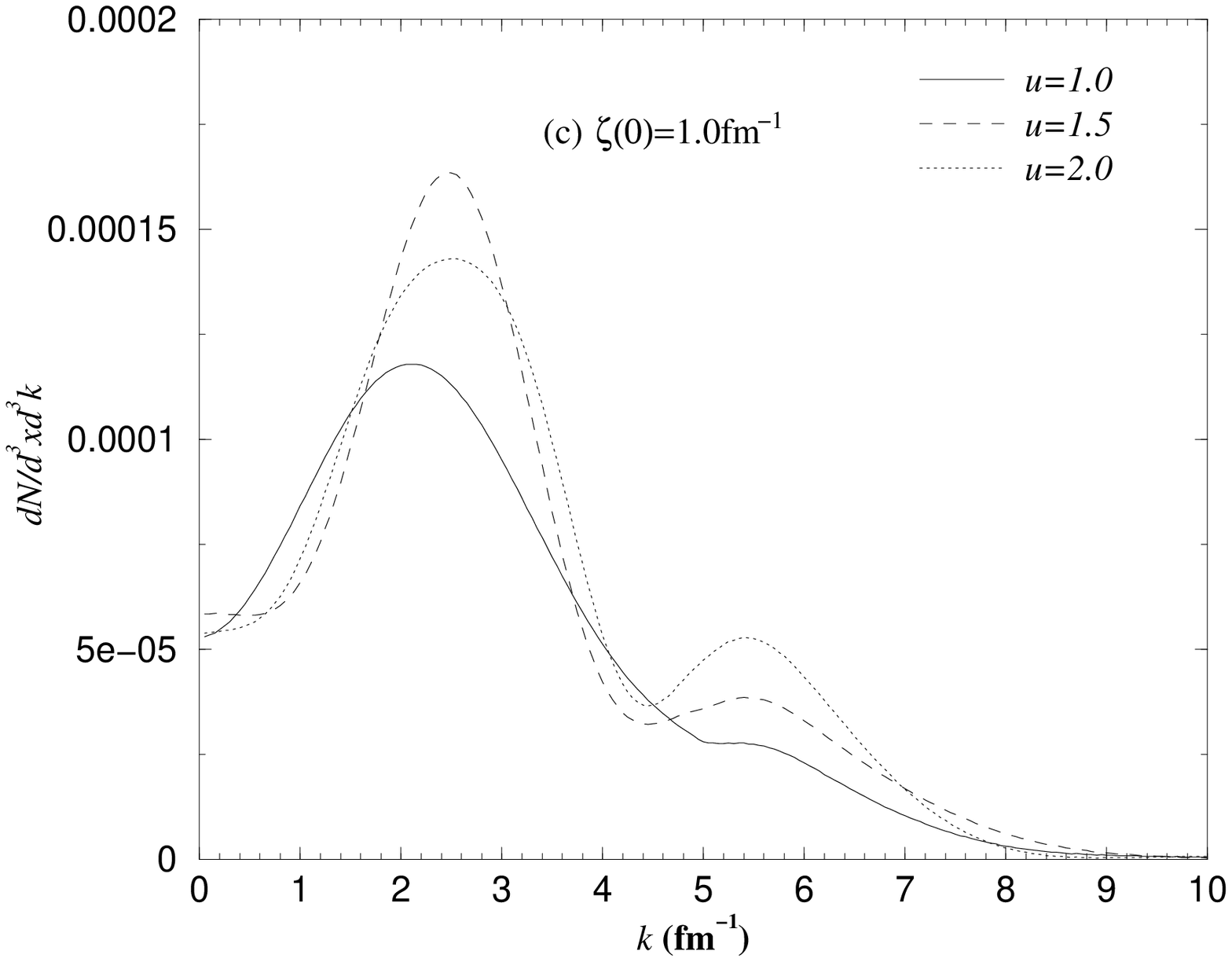}
\caption{ As in Fig.~1 for $\zeta(0)= 1.0~ \rm{fm}^{-1}$. }
\label{fig2}
\end{center}
\end{figure}

The photon spectrum in the laboratory frame emitted from the
non-equilibrium  DCC domains can be obtained by integrating the
rates~(\ref{conformalrate}) over the  spacetime
history~(\ref{dccplab}). To make the comparison, we
calculate the thermal photon emission from the adiabatic
quark-hadron phase transition in Eq.~(\ref{tplab}) with the photon production
rates for the processes discussed above. To do so, as for the non-equilibrium
DCC case, we assume that the QGP is formed with temperature
$T_i =1.0~\rm{fm}^{-1}$ at the formation time $\tau_i
=1.0~\rm{fm}$. In addition, the cooling law in the QGP
phase, i.e. $\tau =(T/T_i) \tau_i$, gives the time
scale  $ \tau_Q =1.1~\rm{fm}$ when the QGP begins hadronizing and
the subsequent quark-hadron mixed state is formed at the
critical temperature assumed to be $T_c =0.9~\rm{fm}^{-1}$.
One can also estimate the time when the mixed state ends,
$\tau_H = ( s^c_Q/s^c_H )^{1/3}=1.77~\rm{fm}$, where the entropy densities
are evaluated at $T=T_c$. In the
final stage,  the system turns into the hadronic phase starting
at $\tau_H$ till the freeze-out time $\tau_f=5.6~\rm{fm}$ estimated from
the cooling law of the hadronic phase,
$\tau_f=\tau_H ( T_c/T_f )^{(\delta+3)/3}$,
where $\delta \simeq 3.4$~\cite{alam} and the freeze-out temperature is chosen
as $T_f=0.5~\rm{fm}^{-1}$.

\begin{figure}[tbp]
\begin{center}
\leavevmode
\epsfxsize 6in
\epsfbox{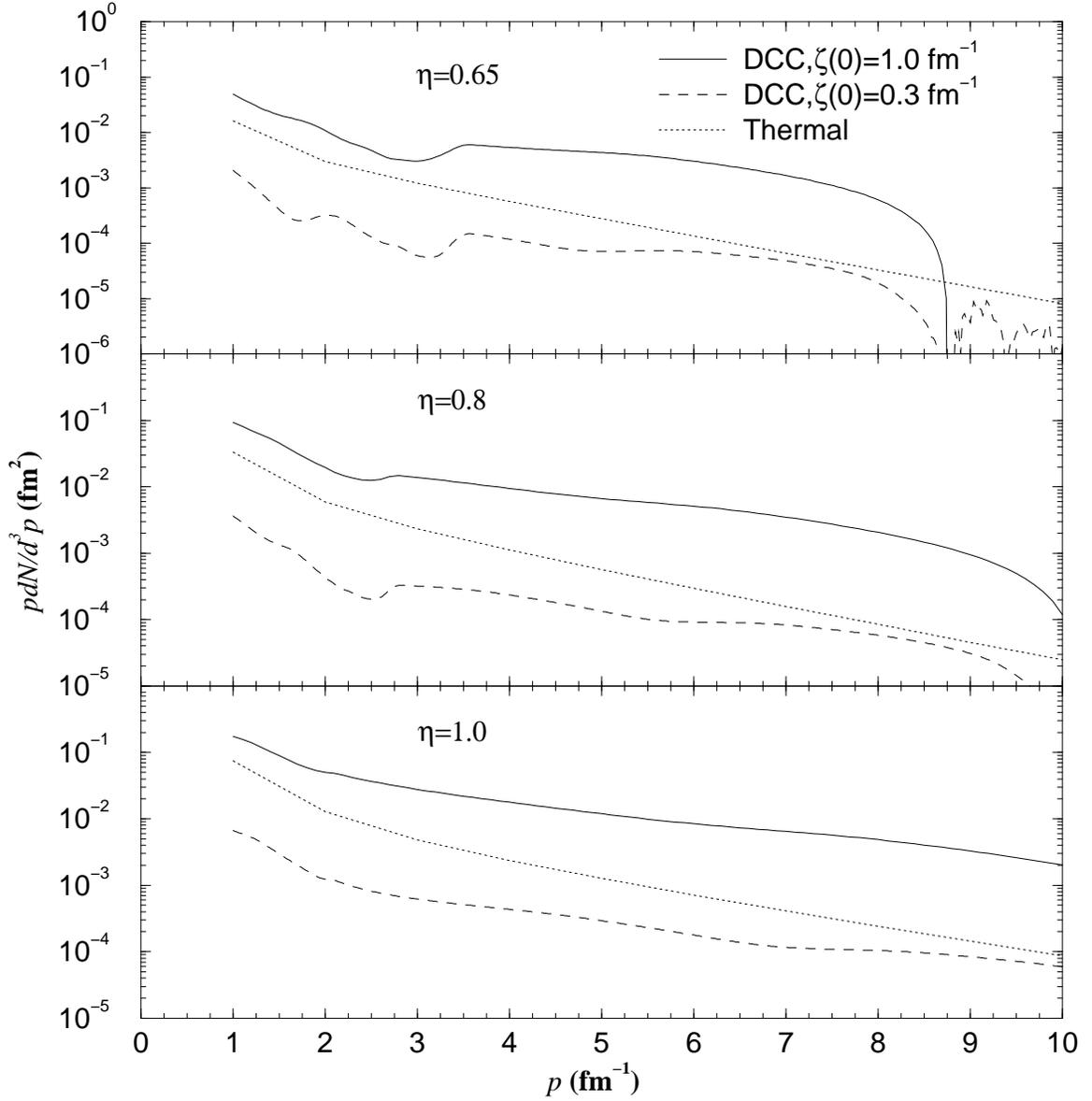}
\caption{Photon energy spectrum measured in laboratory. The solid line denotes
the initial mean field $\zeta(0)= 1.0~ \rm{fm}^{-1}$, while the dashed line
denotes $\zeta(0)= 0.3~ \rm{fm}^{-1}$. The dotted line is the thermal photons
emitted from quark-gluon phase transition. }
\label{fig3}
\end{center}
\end{figure}

Figs.~3(a),~3(b), and~3(c) display the photon emission in the
laboratory frame from the non-equilibrium DCC domains
(solid line for $\zeta(0)= 1.0~ \rm{fm}^{-1}$
and dashed line for $\zeta(0)= 0.3~ \rm{fm}^{-1}$)
and the adiabatic evolution of the QGP phase down to
the hadronic phase (dotted line) for the spacetime
rapidity $\eta =0.65$, $0.8$, and $1.0$ respectively. In all three
cases, our results indicate that the non-equilibrium photon
emission from DCC domains for an initial value of the $ \pi^{0}$
component of an order parameter about $ 1.0~ \rm{fm}^{-1}$ that
undergoes large amplitude oscillations with respect to the minimum
value of the effective action is an order of magnitude larger
than the photon emission from the QGP to the hadronic phase
through an adiabatic first order phase transition. On the
other hand, for an initial value of the $\pi^{0}$ below $ 0.3~
\rm{fm}^{-1}$, due to small amplitude oscillations, the
non-equilibrium DCC photon emission is not so effective  as
compared to the thermal photon emission.
Therefore, we believe that if DCC domains are
indeed formed, the nonlinear dynamics of the $\pi^{0}$
oscillations  can create an enormous amount of non-equilibrium
photons from the DCC domains.
These photon emission enhancement effects could potentially
provide a distinct experimental signature. In addition, the
emitted non-equilibrium photon spectra from DCC domains in a
hydrodynamical expansion shown in Fig.~3
reveal the dramatically different features as compared with the
results in Refs.~\cite{boy2,leeng}, where the authors have considered the
similar non-equilibrium phenomena in the
temperature quenching scenario without convoluting the photon
production rate with the plasma expansion, and shown that
prominent peaks appear in the photon spectrum locating at resonant
momenta of the parametric amplification
determined by the frequency of the $\pi^{0}$ oscillations.
Since the measured emitted photons are chosen with the momentum
along the $ \hat{z}$ direction and the plasma expansion
is along the radial direction, the Doppler effect would cause the
emitted photons either blue-shifted if the photon comes
from the place where the direction of its momentum is  along with
that of the expansion at that place, or red-shifted if the direction of its
momentum is opposite to that of the expansion. As a result,
as we can see in Fig.~3,  the Doppler effect leads to smoothing
out the resonant peaks found in Refs.~\cite{boy2,leeng}.
 In addition,  one can also
estimate the corresponding boost factor arising from
the background expansion effect. From Eq.~(\ref{qp}) and $k=q\tau/\tau_i$, 
one can relate the momentum $k$ in the comformal frame to the momentum $p$
in the laboratory frame by $p\simeq e^\eta k$, where $\eta=0.65$, $0.8$ 
and $1.0$ in Fig.~3.  This
background expansion may  boost  the photon energy, say below the
ultraviolet momentum cutoff of 1 GeV produced from  the low-energy
effective interactions in Eqs.~(\ref{Lpi}--\ref{effpia}), to the energy
as high as the order of $2$ GeV.

In conclusion, we have studied  the production of photons  through the
non-equilibrium relaxation of a
disoriented chiral condensate within which  the chiral order parameter
initially has a non-vanishing expectation value along the $\pi^{0}$ direction,
taking into account the hydrodynamical expansion of the central rapidity
region. The production of non-equilibrium photons driven by the
oscillation of the $\pi^0$ field due to parametric amplification
is modified by the expansion which smoothes out the resonant photon production
in the ''quench'' scenario. However, the resulting photon spectrum exceeds
that for thermal photons from quark-gluon plasma and
hadronic matter for photon energies around $0.2-2\;{\rm GeV}$.
These non-thermal photons can be a potential test of the formation of
disoriented chiral condensates in relativistic heavy-ion-collision experiments.

\vskip 0.3cm

We would like to thank D. Boyanovsky, Hector de Vega, and S.-Y. Wang for their useful discussions.
The work of Y.-Y. Charng, K.-W. Ng, and D.-S. Lee  were supported in part by the
National Science Council, ROC under the grants NSC90-2811-M-001-071,
NSC90-2112-M-001-028, and NSC90-2112-M-259-011 respectively.


\end{document}